\documentclass[preprint,showpacs,preprintnumbers,amsmath,amssymb]{revtex4}

\usepackage{graphicx}
\usepackage{dcolumn}
\usepackage{bm}

\begin{document}


\title{Temperature-doping phase diagram of 
       layered superconductors}

\author{V.M. Loktev}
 \altaffiliation{Electronic address: vloktev@bitp.kiev.ua}
\affiliation{%
Bogolyubov Institute for Theoretical 
        Physics, Metrologichna str. 14-b, Kiev, 03143 Ukraine}%

\author{V. Turkowski}
\altaffiliation{Electronic address: vturk@cfif.ist.utl.pt}
\affiliation{CFIF, Instituto Superior T\' ecnico,
         Av.Rovisco Pais, 1049-001 Lisbon, Portugal}%

\date{April 9, 2003}

\begin{abstract}
The superconducting properties of a layered system are analyzed
for the cases of zero- and non-zero angular momentum of the pairs.
The effective thermodynamic potential for the quasi-2D XY-model
for the gradients of the phase of the order parameter is derived 
from the microscopic superconducting Hamiltonian. The dependence 
of the superconducting critical temperature $T_{c}$ on doping, 
or carrier density, is studied at different values of coupling 
and inter-layer hopping. It is shown that the critical temperature 
$T_{c}$ of the layered system can be lower than the critical 
temperature of the two-dimensional Berezinskii-Kosterlitz-Thouless
transition $T_{BKT}$ at some values of the model parameters, 
contrary to the case when the parameters of the XY-model do not 
depend on the microscopic Hamiltonian parameters. 
\end{abstract}



\maketitle

\section{\label{sec:level1} Introduction}

Theoretical description of the doping dependence of 
the superconducting properties of the
high-temperature superconductors (HTSCs) is one of the most
difficult problem of the modern condensed matter physics.
Generally speaking, the complicated crystal structure of these materials, 
low-dimensional
(quasi-2D) transport properties, the superconducting order parameter
anisotropy, strong correlations
and other properties result in the fact
that many years after the discovery the microscopic
mechanism of HTSC is not understood yet.

During the last years many models which take into account some of the
cuprate properties have been proposed. The doping dependence
of the superconducting properties at $T=0$ in the $s$-wave pairing
channel was studied  
for the 3D case in \cite{Micnas,Randeria,Haussmann} and, particularly
for the quasi-2D case \cite{Gorbar1}. For the 2D case this problem
was studied at $T=0$ ( when a long-range    
superconducting order is still possible in a 2D system 
\cite{Mermin})
for the case of local attraction in 
\cite{Micnas,Randeria,Gorbar2},  
and for the phonon-mediated model \cite{TMF} (for over-review see 
\cite{PhysRep}, for example).
The $d$-wave pairing along with the $s$-wave one at $T=0$
for the case of 
the extended Hubbard model with the next nearest neighbor attraction
was studied in \cite{denHertog,Andrenacci} and for a 2D continuum model
with short-ranger attraction and electron correlations - in the paper 
\cite{Duncan}. Th properties of a model with doping dependent correlation 
length were studied recently in \cite{magnons}.

The 2D $s$-wave pairing at finite temperatures,
when the Berezinskii-Kosterlitz-Thouless (BKT) transition can take 
place in superconducting system, was considered in \cite{Gusynin1,Babaev1} 
for the case of the
model with local attraction
and in
\cite{PhysicaC,JETP} for the case of the electron-phonon pairing. 
The problem of the $s$-wave superconductivity with fluctuating
order parameter phase
in the 3D case was analyzed in \cite{Babaev1,Kopec}.
The effective action for slowly fluctuating $d$-wave
superconducting order parameter for the 2D case was also analyzed in
\cite{Paramekanti1,Benfatto,Sharapov,Paramekanti2}.

However, it is known that the  
long-range order is impossible
in the 1D and 2D systems with an order parameter which has
a continuous symmetry \cite{Mermin}.
Therefore, to get
real phase transition with a long-range order and homogeneous
order parameter
one needs to take
into account the inter-layer coupling $t_{z}$.
The layered superconductivity is much more complicated since
the possibility of the inter-layer fluxon and intra-layer 
vortex phase transitions with corresponding critical temperatures $T_f$
and $T_v$ must be analyzed. 
It was already shown \cite{Hikami,Korshunov,Horovitz1,Horovitz2},
that there is only one phase transition in such a system
with the critical temperature $T_c$ and $T_v<T_c<T_f\simeq 8T_v$.
The critical temperature $T_c$ is equal to $T_v$, or, what
is equivalent, to the temperature $T_{BKT}$
of the 2D BKT phase transition 
at $t_{z}=0$. Then, this temperature value is increasing to the value $T_f$
with the inter-layer hopping $t_{z}$ growth. In the papers
\cite{Hikami,Korshunov,Horovitz1,Horovitz2} the phase order parameter
effective Hamiltonian was studied in the presence of an external
magnetic field and this model was mapped on the quasi-2D
XY-model. The XY-model parameters $J_{\parallel}$ and $J_{\perp}$
were considered as phenomenological constants. It was shown 
that $T_{BKT}=\frac{\pi}{2} J_{\parallel}$ and $T_{f}\simeq 8T_{BKT}$.

In this paper we derive the effective XY-Hamiltonian
from the initial Hamiltonian for the layered system of attracting
fermions.  In this case the parameters $J_{\parallel}$ and $J_{\perp}$
depend on the bare parameters - charge carrier
density, coupling, pair angular momentum, temperature
and the inter-layer hopping.
As it will be shown below, this leads to the non-trivial 
dependence of the superconducting critical temperature $T_{c}$ on the
model parameters. In particular, in general this temperature
is different from the critical temperature of the 2D BKT transition
and $T_{c} <T_{BKT}$ at some values of the model parameters,
contrary to the results for the case when parameters $J_{\parallel}$
and $J_{\perp}$ don't depend on the parameters of the microscopic
Hamiltonian, and when the relation $T_{c} <T_{BKT}$ always holds
at $J_{\perp}>0$ \cite{Hikami,Korshunov,Horovitz1,Horovitz2}.

\section{\label{sec:level2} The model and the thermodynamic
potential}

The model Hamiltonian for a layered superconducting system can be written as
\begin{eqnarray}
H(\tau )=\sum_{\sigma ,j}\int d^{2}r
\psi_{j\sigma}^{\dagger}(\tau ,{\bf r})
\left[-\frac{\vec{\nabla}^{2}}{2m}+2t_{z}-\mu \right]
\psi_{j\sigma}(\tau ,{\bf r})
-\sum_{\sigma ,j_{1},j_{2}}t_{mn}\int d^{2}r
\psi_{j_{1}\sigma}^{\dagger}(\tau ,{\bf r})
\psi_{j_{2}\sigma}(\tau ,{\bf r})
\nonumber \\
-\frac{1}{2}\sum_{\sigma ,j}\int d^{2}r_{1}d^{2}r_{2}
\psi_{j\sigma}^{\dagger}(\tau ,{\bf r}_{2})
\psi_{j{\bar \sigma}}^{\dagger}(\tau ,{\bf r}_{1})
V({\bf r}_{1},{\bf r}_{2})
\psi_{j{\bar \sigma}}(\tau ,{\bf r}_{1})
\psi_{j\sigma}(\tau ,{\bf r}_{2}),
\label{Hamiltonian}   
\end{eqnarray}
where $\psi_{j\sigma}(\tau ,{\bf r})$ is a fermi-field
in the with mass $m$ and spin $\sigma =\uparrow ,\downarrow$, 
$\tau$ is an imaginary time and $j,{\bf r}$ are layer 
number and intra-layer
coordinate, correspondingly; 
$t_{j_{1}j_{2}}=t_{z}(\delta_{j_{2},j_{1}+1}+\delta_{j_{2},j_{1}-1})$ 
corresponds to the nearest inter-plane hopping.
The free fermion dispersion relation in the momentum
space has the following form 
\begin{equation}
\xi ({\bf k},k_{z})=\frac{{\bf k}^{2}}{2m}+2t_{z}-2t_{z}\cos (ak_{z})-\mu ,
\label{xi}
\end{equation}
where ${\bf k}$ is a 2D wave vector with a bandwidth $W$,
and $k_{z}$ is the momentum in the inter-layer ($z$) direction,
it changes in the interval $[0,2\pi /a]$, where $a$ is the
inter-layer spacing; $\mu$ is the chemical potential
of the system.
In Eq.(\ref{Hamiltonian}) interaction $V({\bf r}_{1},{\bf r}_{2})$ describes
a non-retarded in-plane fermion attraction.

The partition function of the system is 
\begin{equation}
Z=\int D\psi^{\dagger}D\psi e^{-S}
\label{Z}
\end{equation}
with the action
\begin{equation}
S=\int_{0}^{\beta}d\tau 
\left[ 
\sum_{\sigma ,j}\int d^{2}r
\psi_{j\sigma}^{\dagger}(\tau ,{\bf r})
\partial_{\tau}
\psi_{j\sigma}(\tau ,{\bf r})+H(\tau)
\right] .
\label{S}
\end{equation}

To study the superconducting properties of the system
with an arbitrary pairing symmetry
the Hubbard-Stratonovich transformation with bilocal fields
$\phi_{j} (\tau ,{\bf r}_{1},{\bf r}_{2})$ and 
$\phi_{j}^{\dagger}(\tau ,{\bf r}_{1},{\bf r}_{2})$
can be applied \cite{Kleinert1}:
\begin{eqnarray}
\exp \left[ \psi_{j\uparrow}^{\dagger}(\tau ,{\bf r}_{2})
\psi_{j\downarrow}^{\dagger}(\tau ,{\bf r}_{1})
V({\bf r}_{1},{\bf r}_{2})
\psi_{j\downarrow}(\tau ,{\bf r}_{1})
\psi_{j\uparrow}(\tau ,{\bf r}_{2})\right]
=\int D\phi^{\dagger}D\phi
\exp\left[ -\int_{0}^{\beta}d\tau\sum_{j}\int d^{2}r_{1}d^{2}r_{2}
\right.
\nonumber \\
\left.
\times
\left(
\frac{|\phi_{j} (\tau ,{\bf r}_{1},{\bf r}_{2})|^{2}}
{V({\bf r}_{1},{\bf r}_{2})}
-\phi_{j}^{\dagger}(\tau ,{\bf r}_{1},{\bf r}_{2})
\psi_{j{\downarrow}}(\tau ,{\bf r}_{1})
\psi_{j\uparrow}(\tau ,{\bf r}_{2})
-\psi_{j{\uparrow}}^{\dagger}(\tau ,{\bf r}_{1})
\psi_{j\downarrow}^{\dagger}(\tau ,{\bf r}_{2})
\phi_{j}(\tau ,{\bf r}_{1},{\bf r}_{2})
\right)\right] .
\label{HS}
\end{eqnarray}
Let us introduce the Nambu spinor
$$
\Psi_{j} (\tau ,{\bf r}) = 
\left(
\begin{array}{c}
\psi_{j\uparrow} (\tau ,{\bf r}) \\
\psi_{j\downarrow}^{\dagger} (\tau ,{\bf r})  
\end{array}\right)\ ,
\Psi_{j}^{\dagger} (\tau ,{\bf r}) = 
\left(
\psi_{j\uparrow}^{\dagger} (\tau ,{\bf r}),
\psi_{j\downarrow} (\tau ,{\bf r})  
\right) .
$$
In this case the partition function can be written as
\begin{equation}
Z=\int D\psi^{\dagger}D\psi D\phi^{\dagger}D\phi 
e^{-\bar{S}(\psi^{\dagger},\psi ,\phi^{\dagger} ,\phi)},
\label{Z2}
\end{equation}
where
\begin{eqnarray}
\bar{S}(\psi^{\dagger},\psi ,\phi^{\dagger} ,\phi)
=\int_{0}^{\beta}d\tau \sum_{j_{1},j_{2}}
\int d^{2}r_{1}\int d^{2}r_{2}
\{ \delta_{j_{1}j_{2}}\frac{|\phi_{j_{1}} (\tau ,{\bf r}_{1},{\bf r}_{2})|^{2}}
{V({\bf r}_{1},{\bf r}_{2})}
\nonumber \\ 
-\delta_{j_{1}j_{2}}
\Psi_{j_{1}}^{\dagger}(\tau ,{\bf r}_{1})
\left[-\partial_{\tau}-\tau_{z}(\frac{\vec{\nabla}_{{\bf r}_{1}}^{2}}{2m}+2t_{z}-\mu)\right]
\Psi_{j_{1}}(\tau ,{\bf r}_{2})\delta ({\bf r}_{1}-{\bf r}_{2})
\nonumber \\ 
+
t_{j_{1}j_{2}}\Psi_{j_{1}}^{\dagger}(\tau ,{\bf r}_{1})\tau_{z}
\Psi_{j_{2}}(\tau ,{\bf r}_{2})\delta ({\bf r}_{1}-{\bf r}_{2})
-
\delta_{j_{1}j_{2}}\phi_{j_{1}}^{\dagger}(\tau ,{\bf r}_{1},{\bf r}_{2})
\Psi_{j_{1}}^{\dagger}(\tau ,{\bf r}_{1})
\tau_{-}
\Psi_{j_{1}}(\tau ,{\bf r}_{2})
\nonumber \\
-\delta_{j_{1}j_{2}}\Psi_{n}^{\dagger}(\tau ,{\bf r}_{1})
\tau_{+}
\Psi_{j_{1}}(\tau ,{\bf r}_{2})
\phi_{j_{1}}(\tau ,{\bf r}_{1},{\bf r}_{2})\} ,
\label{Sb}
\end{eqnarray}
where $\tau_{\pm}=\frac{1}{2}(\tau_{x}\pm \tau_{y})$ are the
Pauli matrices.

In order to study the fluctuations of the order parameter phase
and to map the corresponding superconducting effective action on the
quasi-2D $XY$ model,
it is convenient
to make decomposition of $\psi_{\sigma,j}(\tau , {\bf r})$
$\psi_{\sigma,j}^{\dagger}(\tau , {\bf r})$ on
their modulus $\chi_{\sigma,j}(\tau , {\bf r})$ 
and phase $\theta_{j}(\tau , {\bf r})$, which as it will
be shown below is proportional to the order parameter phase: 
$$
\psi_{\sigma,j}(\tau , {\bf r})=
\chi_{\sigma,j}(\tau , {\bf r})e^{i\theta_{j}(\tau , {\bf r})/2},
$$
$$
\psi_{\sigma,j}^{\dagger}(\tau , {\bf r})=
\chi_{\sigma,j}^{\dagger}(\tau , {\bf r})e^{-i\theta_{j}(\tau , {\bf r})/2}.
$$
In this case the Nambu operators are
\begin{eqnarray}
\Psi_{j}(\tau ,{\bf r})=e^{i\tau_{z}\theta_{j} (\tau ,{\bf r})/2}
\Upsilon_{j}(\tau ,{\bf r}),
\nonumber \\
\Psi_{j}^{\dagger}(\tau ,{\bf r})=\Upsilon_{j}^{\dagger}(\tau ,{\bf r})
e^{-i\tau_{z}\theta_{j} (\tau ,{\bf r})/2},
\label{Nambu2}
\end{eqnarray}
where $\Upsilon_{j}(\tau ,{\bf r})$ and 
$\Upsilon_{j}^{\dagger}(\tau ,{\bf r})$
are ``neutral'' Nambu spinor operators:
$$
\Upsilon_{j} (\tau ,{\bf r}) = 
\left(
\begin{array}{c}
\chi_{j\uparrow} (\tau ,{\bf r}) \\
\chi_{j\downarrow}^{\dagger} (\tau ,{\bf r})  
\end{array}\right)\ ,
\Upsilon_{j}^{\dagger} (\tau ,{\bf r}) = 
\left(
\chi_{j\uparrow}^{\dagger} (\tau ,{\bf r}),
\chi_{j\downarrow} (\tau ,{\bf r})  
\right) .
$$
The order parameter can be expressed as 
$$
\phi_{j}(\tau ,{\bf r}_{1},{\bf r}_{2})=
\Delta (\tau ,{\bf r}_{1},{\bf r}_{2})
e^{i\theta_{j}(\tau ,{\bf r}_{1},{\bf r}_{2})}
$$
$$
\phi_{j}^{\dagger}(\tau ,{\bf r}_{1},{\bf r}_{2})=
\Delta (\tau ,{\bf r}_{1},{\bf r}_{2})
e^{-i\theta_{j}(\tau ,{\bf r}_{1},{\bf r}_{2})},
$$
where we assume that the modulus of the order parameter
$\Delta (\tau ,{\bf r}_{1},{\bf r}_{2})$ does not
depend on the layer index.
It is also natural to assume that
\begin{equation} 
\phi_{j}(\tau ,{\bf r}_{1},{\bf r}_{2})\simeq
\Delta (\tau ,{\bf r})
e^{i\theta_{j}(\tau ,{\bf R})},
\label{appr}
\end{equation}
where 
${\bf r}={\bf r}_{1}-{\bf r}_{2}$ 
and 
${\bf R}=({\bf r}_{1}+{\bf r}_{2})/2$
are the relative and
the center of mass coordinates, correspondingly
\cite{Ao,Sharapov}.
The relation (\ref{appr}) means that the dynamics of the Cooper pairs
is described by the order parameter modulus the symmetry of which 
depends, generally speaking, on the relative
pair coordinate and the motion of the superconducting condensate
is described by the order parameter phase, which changes
slowly with the distance and can be described by center of 
mass coordinate.
In this case it is easy to obtain
\begin{eqnarray}
\phi_{j}^{\dagger}(\tau ,{\bf r}_{1},{\bf r}_{2})
\Psi_{j}^{\dagger}(\tau ,{\bf r}_{1})\tau_{-}
\Psi_{j}(\tau ,{\bf r}_{2})
+
\Psi_{j}^{\dagger}(\tau ,{\bf r}_{1})\tau_{+}
\Psi_{j}(\tau ,{\bf r}_{2})
\phi_{j}(\tau ,{\bf r}_{1},{\bf r}_{2})
\nonumber \\
\simeq
\Delta (\tau ,{\bf r})
\Upsilon_{j}^{\dagger}(\tau ,{\bf r}_{1})\tau_{x}
\Upsilon_{j}(\tau ,{\bf r}_{2})
\label{appr2}
\end{eqnarray}
Substituting (\ref{Nambu2}), (\ref{appr}) and (\ref{appr2}) 
into the expression
for the partition function (\ref{Z2}) it is easy to get
$$
Z=\int\Delta D\Delta D\theta e^{-\beta\Omega (\Delta , \theta)},
$$
where the thermodynamic potential is
$$
\beta\Omega (\Delta , \theta)=
\int_{0}^{\beta}d\tau\int d^{2}r
\frac{N\Delta (\tau ,{\bf r})^{2}}
{V({\bf r})}-TrlnG^{-1}, 
$$
$N$ is number of the layers. 
The Nambu spinor Green function $G$ can be expressed as
$$
G^{-1}={\cal G}^{-1}-\Sigma ,
$$
where ${\cal G}^{-1}$ is a part of the inverse Green's function
which does not depend on the order parameter phase:
$$
{\cal G}_{j_{1}j_{2}}^{-1}(\tau_{1},\tau_{2},{\bf r}_{1},{\bf r}_{2})=
\langle \tau_{1},{\bf r}_{1},j_{1}|
{\cal G}^{-1}
|\tau_{2},{\bf r}_{2},j_{2}\rangle 
$$
$$
=\delta_{j_{1}j_{2}}\delta ({\bf r}_{1}-{\bf r}_{2})
\delta (\tau_{1}-\tau_{2})
\left[-\partial_{\tau_{1}}-\tau_{z}\left(-\frac{{\bf \nabla}_{{\bf r}_{1}}^{2}}{2m}
+2t-\mu 
\right)\right]
$$
$$
-\delta_{j_{2},j_{1}\pm 1}\delta ({\bf r}_{1}-{\bf r}_{2})
\delta (\tau_{1}-\tau_{2})\tau_{z} t_{z}
+\delta_{j_{1}j_{2}}\tau_{x}\Delta (\tau_{1}-\tau_{2},{\bf r}_{1}-{\bf r}_{2}).
$$
The self-energy $\Sigma$ is the sum of the parts
which come from the in-plane and inter-plane order parameter
phase phase interaction $\Sigma^{\parallel}$ and $\Sigma^{\perp}$,
respectively:
$$
\Sigma = \Sigma^{\parallel}+ \Sigma^{\perp},
$$
where
$$
\Sigma^{\parallel}_{j_{1}j_{2}}
(\tau_{1},\tau_{2},{\bf r}_{1},{\bf r}_{2})=
\langle \tau_{1},{\bf r}_{1},j_{1}|
\Sigma^{\parallel}
|\tau_{2},{\bf r}_{2},j_{2}\rangle 
$$
$$
=\delta_{j_{1}j_{2}}\delta ({\bf r}_{1}-{\bf r}_{2})
\delta (\tau_{1}-\tau_{2})
\left[
\frac{i\tau_{z}}{2}\partial_{\tau_{1}}\theta_{j_{1}}(\tau_{1},{\bf r}_{1})
 -\frac{i}{4m}{\bf \nabla}_{{\bf r}_{1}}^{2}
\theta_{j{1}}(\tau_{1},{\bf r}_{1})
\right.
$$
$$
\left.
 +\frac{\tau_{z}}{8m}({\bf \nabla}_{{\bf r}_{1}}
\theta_{j_{1}}(\tau_{1},{\bf r}_{1}))^{2}
-\frac{i}{2m}{\bf \nabla}_{{\bf r}_{1}}\theta_{j_{1}}(\tau_{1},{\bf r}_{1})
{\bf \nabla}_{{\bf r}_{1}}
\right]
$$
and
$$
\Sigma^{\perp}_{j_{1}j_{2}}
(\tau_{1},\tau_{2},{\bf r}_{1},{\bf r}_{2})=
\langle \tau_{1},{\bf r}_{1},j_{1}|
\Sigma^{\perp}
|\tau_{2},{\bf r}_{2},j_{2}\rangle 
$$
$$
=-\delta_{j_{2},j_{1}\pm 1}\delta ({\bf r}_{1}-{\bf r}_{2})
\delta (\tau_{1}-\tau_{2})\tau_{z}t_{z}
(1-\exp[-i\tau_{z}(\theta_{j_{1}}(\tau_{1},{\bf r}_{1})-
\theta_{j_{2}}(\tau_{2},{\bf r}_{2}).
)])
$$

The potential term of the thermodynamic potential is
$$
\beta\Omega_{pot} (\Delta )=
\int_{0}^{\beta}d\tau \int d^{2}r
\frac{N\Delta (\tau ,{\bf r})^{2}}
{V({\bf r})}-Trln{\cal G}^{-1}. 
$$
and the kinetic term can be expanded in powers of the self-energy
$\Sigma$:
\begin{equation}
\beta\Omega_{kin} (\Delta , \theta)=
Tr\sum_{n=1}^{\infty}\frac{1}{n}({\cal G}\Sigma)^{n}.
\label{Omegakin}
\end{equation}

\section{\label{sec:level2} The BKT transition in the 2D case}

Let us begin with the case when there is no inter-plane coupling:
$t_{z}=0$. In this case the behavior in each plane is independent
and the system undergoes the BKT
transition. Let us assume that the order parameter phase
fluctuations are small. In this case to get
the thermodynamic potential up to the second order in ${\bf\nabla}\theta$
we neglect all the terms in (\ref{Omegakin}), except $n=1,2$. 
Also we neglect the time dependence of $\theta$ and the second derivative
${\bf \nabla}^{2}\theta$.
The effective potential in this case has the following structure 
(see, for example \cite{PhysRep}):
\begin{equation}
\Omega (\Delta , \theta)=\Omega_{pot} (\Delta )+
\frac{J_{\parallel}}{2}\int d^{2}r ({\bf\nabla}\theta )^{2},
\label{Omega2D}
\end{equation}
where
\begin{eqnarray}
J_{\parallel}=\int \frac{d^{2}kdk_{z}}{(2\pi )^{3}}
\left(
\frac{n_{f}({\bf k})}{4m}
-
\frac{1}{16m^{2}}\frac{1}{T}
\frac{{\bf k}^{2}}{\cosh^{2}(\sqrt{\xi ({\bf k})^{2}+\Delta ({\bf k})^{2}}/2T)}
\right)
,
\label{Jpar}
\end{eqnarray}
and the momentum distribution function $n_{f}({\bf k})$ is
\begin{equation}
n_{f}({\bf k})=1-
\tanh\left(\frac{\sqrt{\xi ({\bf k})^{2}+\Delta ({\bf k})^{2}}}{2T}\right)
\frac{\xi ({\bf k})}{\sqrt{\xi ({\bf k})^{2}+\Delta ({\bf k})^{2}}}.
\label{MDF}
\end{equation}
Free fermion spectrum $\xi ({\bf k})$ in (\ref{Jpar}) and (\ref{MDF})
is defined by (\ref{xi}) at $t_{z}=0$ in this case.

The minimization of the effective potential (\ref{Omega2D}) 
at ${\bf\nabla}\theta =0$
with respect
to the superconducting order parameter $\Delta ({\bf k})$ leads
to the standard gap equation:
\begin{eqnarray}
\Delta ({\bf p})=\int \frac{d^{2}kdk_{z}}{(2\pi )^{3}}
\frac{\Delta ({\bf k})}{2\sqrt{\xi ({\bf k})^{2}+\Delta ({\bf k})^{2}}}
\tanh\left(\frac{\sqrt{\xi ({\bf k})^{2}+\Delta ({\bf k})^{2}}}{2T}
\right) V({\bf p},{\bf k}).
\label{gap}
\end{eqnarray}

The minimization of the effective potential at ${\bf\nabla}\theta =0$
with respect
to the chemical potential $\delta \Omega_{pot}/\delta \mu =-\upsilon n_{f}$ 
($\upsilon$ is the volume of the system)
gives the equation which connects $\mu$ and the particle
density $n_{f}$ in the system, or the 2D Fermi energy
$e_{F}=\pi n_{f}/m$:
\begin{equation}
n_{f}=\int \frac{d^{2}kdk_{z}}{(2\pi )^{3}}n_{f}({\bf k}),
\label{number}
\end{equation}
where the momentum distribution function $n_{f}({\bf k})$
is defined in (\ref{MDF}).

To search the solutions with different angular momenta $l$ of the pairs,
we assume that the interaction potential has the following form:
\begin{equation}
V({\bf p},{\bf k})=V\cos (l\varphi_{\bf p})\cos (l\varphi_{\bf k}).
\label{Vpk}
\end{equation}
Below we shall use dimensionless coupling parameter $G=mV/(2\pi )$
for the numerical calculations. 

In the case of the interaction (\ref{Vpk}) the gap depends 
only on the momentum direction: 
$$
\Delta ({\bf p})=\Delta_{l}\cos (l\varphi_{\bf p}),
$$
where $\Delta_{l}$ is the amplitude of the superconducting gap
in the case of the pair angular momentum equal to $l$.
The solution of the gap equation together with the number equation
at $\Delta_{l}=0$ give the critical temperature of the 
mean-field superconducting
transition $T_{\Delta}\equiv T_{c}^{MF}$ 
on the charge carrier density $n_{f}$.
The solution of the equation 
\begin{equation}
T=\frac{\pi}{2}J_{\parallel}(\Delta_{l},\mu , T)
\label{TBKT}
\end{equation}
together with the gap equation and the number equation give
the dependence of the critical temperature 
of the BKT-transition on the
charge carrier density $n_{f}$. Equation (\ref{TBKT}) is obtained
by mapping (\ref{Omega2D}) on the corresponding
thermodynamic potential of the 2D spin XY-model.
 
As it follows from   
the system (\ref{gap}), (\ref{number}) and (\ref{TBKT}),
the solution for the $T_{\Delta}$ and $T_{BKT}$
do not depend on $l$ when $l\not= 0$ for the case of the simple interaction
potential (\ref{Vpk}).
This follows from the fact that the $l$-dependence of the
integral is only as $\cos^{2}(l\varphi)$ and from the identity
$$
\int_{0}^{2\pi}\frac{d\varphi}{(2\pi )}F\left[ \cos^{2}(l\varphi)\right]
=\int_{0}^{2\pi}\frac{d\varphi}{(2\pi )}F\left[ \cos^{2}(\varphi)\right] ,
$$
where $F\left[ \cos^{2}(l\varphi)\right]$ is an arbitrary function
without singularities, and $l$ is an arbitrary
non-zero integer number. Therefore it is necessary to analyze the
solutions with $l=0$ and $l=1$.

The phase diagram of the system in the 2D case is presented in the
Fig.1. 
The temperature $T_{\Delta}$ is much higher in the $s$-channel.
However, $T_{BKT}\simeq e_{F}/8$ in both channels
at small carrier density. This result can be easily
obtained analytically from (\ref{Jpar}) and (\ref{TBKT}).

The doping dependence of the $T_{BKT}$ in the cases of $l=0$
and $l\neq 0$ is presented in Figs. 2 and 3, correspondingly.
The relation $T_{BKT}\simeq e_{F}/8$ holds
up to higher values of the carrier density in the $s$-channel
at fixed value of coupling. It means that the local pairs
are bounded tighter in this case.

\begin{figure}[h]
\centering{
\includegraphics[width=8cm,angle=270]{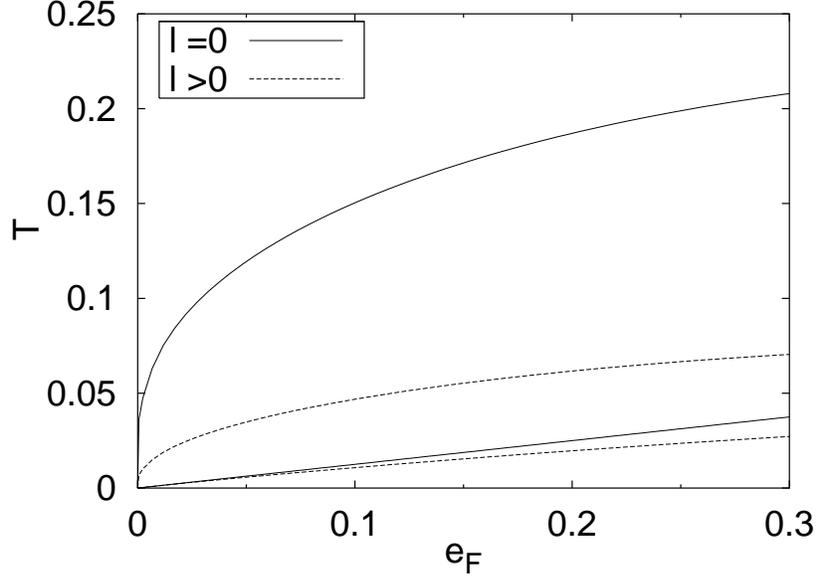}}
\caption{Phase diagram of the 2D system in different
pairing channels for the coupling parameter $G=1$.
The solid lines are $T_{\Delta}$
(the upper curve) and $T_{BKT}$ for the $s$-wave pairing channel.
 The dashed lines are the corresponding curves for the case $l\not= 0$.
Here and below all quantities are normalized on the 2D free electron
bandwidth $W$.
}  
\label{fig:1}
\end{figure}

\begin{figure}[h]
\centering{
\includegraphics[width=8cm,angle=270]{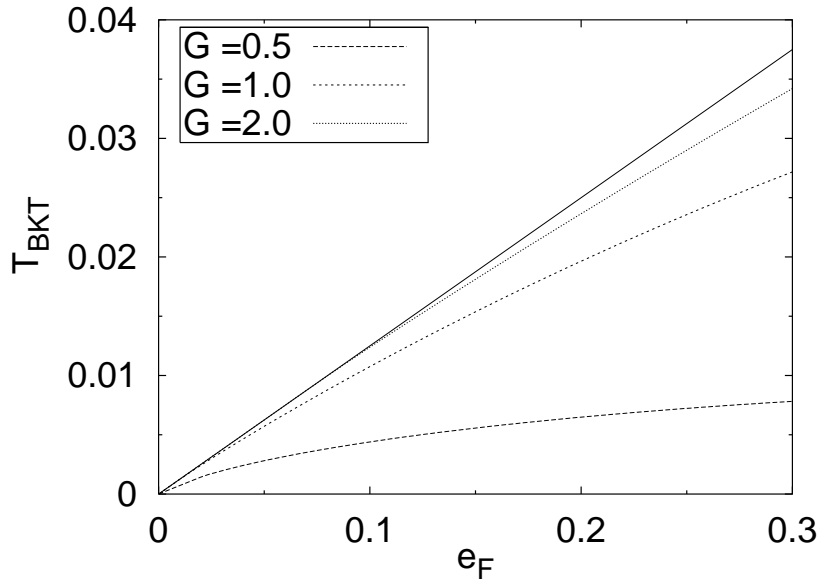}}
\caption{The doping dependence of $T_{BKT}$
at $l\not= 0$ and different coupling parameters:
$G=0.5$ (dash-dotted line), $G=1.0$ (dotted line) and
$G=2.0$ (dashed line). The solid line is the function
$T_{BKT}=e_{F}/8$.} 
\label{fig:2}
\end{figure}

\begin{figure}[h]
\centering{
\includegraphics[width=8cm,angle=270]{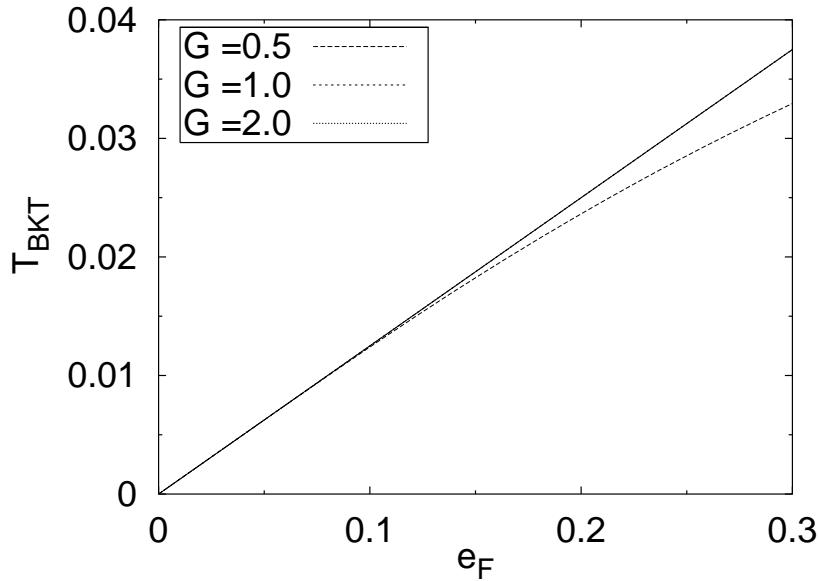}}
\caption{The same as in the previous Figure
for the case $l=0$. The lines for
$G=1.0$ and for $G=2.0$ practically coincide
with $T_{BKT}=e_{F}/8$.} 
\label{fig:3}
\end{figure}

\section{\label{sec:level1} Transition in the case of coupled
layers}

Let us consider a system of coupled layers. The self-energy,
proportional to the inter-layer coupling can be written as
$$  
\Sigma^{\perp}=t_{z}\tau_{z}\Sigma^{\perp}_{1}+t_{z}\Sigma^{\perp}_{2},
$$
where
$$
\Sigma^{\perp}_{1\ \ j_{1}j_{2}}
(\tau_{1},\tau_{2},{\bf r}_{1},{\bf r}_{2})=
\langle \tau_{1},{\bf r}_{1},j_{1}|
\Sigma^{\perp}_{1}
|\tau_{2},{\bf r}_{2},j_{2}\rangle 
$$
$$
=-\delta_{j_{2},j_{1}\pm 1}\delta ({\bf r}_{1}-{\bf r}_{2})
\delta (\tau_{1}-\tau_{2})
\cos (\theta_{j_{1}}(\tau_{1},{\bf r}_{1})-
\theta_{j_{2}}(\tau_{2},{\bf r}_{2})),
$$
$$
\Sigma^{\perp}_{2\ \ j_{1}j_{2}}
(\tau_{1},\tau_{2},{\bf r}_{1},{\bf r}_{2})=
\langle \tau_{1},{\bf r}_{1},j_{1}|
\Sigma^{\perp}_{2}
|\tau_{2},{\bf r}_{2},j_{2}\rangle 
$$
$$
=\delta_{j_{2},j_{1}\pm 1}\delta ({\bf r}_{1}-{\bf r}_{2})
\delta (\tau_{1}-\tau_{2})
\sin (\theta_{j_{1}}(\tau_{1},{\bf r}_{1})-
\theta_{j_{2}}(\tau_{2},{\bf r}_{2})).
$$
Similarly to the 2D case, we assume that the phase of the order parameter
changes slowly in the inter-layer direction.
Therefore, the thermodynamic potential can be calculated
up to the second order in $(\theta_{j}-\theta_{j\pm 1})$:
\begin{equation}
\Omega_{kin}^{\perp}=
t_{z}TTr({\cal G}\tau_{z}\Sigma^{\perp}_{1})+
\frac{t_{z}^{2}}{2}TTr({\cal G}\Sigma^{\perp}_{2}
{\cal G}\Sigma^{\perp}_{2}). 
\label{Omegakinperp}
\end{equation}
The terms proportional to 
$\Sigma^{\parallel}\Sigma^{\perp}$
and $\Sigma^{\perp}_{1}\Sigma^{\perp}_{2}$
are zero due to reflection symmetry in $z$-direction.

To map the system on the quasi-2D XY-model with
the nearest neighbor interaction we need
to obtain 
\begin{equation}
\Omega_{kin}=
\frac{J_{\parallel}}{2}\sum_{j}\int d^{2}r ({\bf\nabla}\theta_{j})^{2}+
J_{z}\sum_{j}(1-\cos (\theta_{j}-\theta_{j-1})).
\label{Omegakin}
\end{equation}
This dependence comes from the first term in (\ref{Omegakinperp}).
The second term in (\ref{Omegakinperp}) is proportional to 
$\sin (\theta_{j}-\theta_{j\pm1})\sin (\theta_{j}-\theta_{j\pm1})$,
what is equivalent to the XY-model with the next nearest
neighbor and next next nearest neighbor interactions.
Therefore we neglect this term since it is of a higher order ($\sim t_{z}^{2}$)
on the inter-layer hopping with respect to the first term 
(which is $\sim t_{z}$).
However, if the coupling $t_{z}$ is not small this term can lead
for important physical consequences (see for example an analysis for the
2D case \cite{Kim}).
Thus, the parameter $J_{z}$ is
\begin{equation}
J_{z}=t_{z}\int \frac{d^{2}kdk_{z}}{(2\pi )^{3}}
n_{f}({\bf k})\cos (ak_{z}).
\label{Jperp}
\end{equation}

Now we have obtained the kinetic part of the thermodynamic potential 
$\Omega_{kin}$ in the case of slowly fluctuating phase of the order parameter.
This function is given by (\ref{Omegakin}), where
the parameters $J_{\parallel}$ and $J_{z}$ are given by
(\ref{Jpar}) and (\ref{Jperp}). Similarly to (\ref{Jperp}),
an additional integration over $k_{z}$ must be performed in (\ref{Jpar}).

The effective action (\ref{Omegakinperp}) was studied
in \cite{Hikami,Korshunov,Horovitz1,Horovitz2} in the case
when the parameters $J_{\parallel}$ and $J_{z}$ where considered independent
on the fermion Hamiltonian parameters. 
It was shown \cite{Korshunov}, 
that there is only one phase transition 
in such a system at $T_{c}$ which is bigger
than the temperature of the BKT transition in the case
of non-coupled layers $T_{BKT}=(\pi /2)J_{\parallel}$.
In the case of small coupling $T_{c}\simeq T_{BKT}$
and when $t_{z}$ is increasing to the inter-plane hopping value,
$T_{c}$ is approaching to the value $T_{BKT}=4\pi J_{\parallel}\simeq
8T_{BKT}=T_{f}$ of the fluxon transition, when the inter-layer
order starts to take place.

More precisely, the following expression for the
 effective free energy was considered
\begin{eqnarray}
{\cal F}=\frac{1}{8\pi}\int d^{2}rdz
\{ 
({\bf \nabla}\times {\bf A})^{2}
+\frac{1}{\lambda_{e}}\sum_{j}[\frac{\phi_{0}}{2\pi}
{\bf \nabla}\theta_{j}({\bf r})-{\bf A}({\bf r},z)
]^{2}\delta (z-jd)
\}
\nonumber \\
-\frac{J_{z}}{\xi_{0}^{2}}\int d^{2}r
\cos [
\theta_{j}({\bf r})-\theta_{j-1}({\bf r})-
\frac{2\pi}{\phi_{0}}\int_{(j-1)d}^{jd}
A_{z}({\bf r},z')dz']
-E_{c}\sum_{j,{\bf r}}s_{j}^{2}({\bf r}),
\label{freeenergy}
\end{eqnarray}
where ${\bf A}({\bf r},z)$ is the vector potential,
$\phi_{0}=hc/2e$ is the flux quantum, $E_{c}$ is the loss
of the condensation energy in a volume $\xi_{0}^{2}d$,
$\xi_{0}$ is the in-plane correlation length, $d_{0}$
is the thickness of each layer, and $d(>d_{0})$ is 
the inter-layer distance.
The field $s_{j}({\bf r})$ describes vorticity of the lattice,
$s_{j}({\bf r})=1$ if the vortex is present at the point,
and $s_{j}({\bf r})=0$, otherwise.
The length scale $\lambda_{e}$ is connected with the London 
in-plane penetration length $\lambda_{L}$ as 
$\lambda_{e}=\lambda_{L}^{2}/d_{0}$.
It was shown by a renormalization group study \cite{Horovitz1,Horovitz2}
that in a physical case $\lambda_{e}\gg d_{0}$
the self-consistent equation which describes the dependence of the critical
temperature $T_{c}$ on the free energy parameters (\ref{freeenergy})
has the form
\begin{equation}
T_{c}\simeq 
\frac{\tau [E_{c}+(\tau /8)ln(T_{c}/J_{z})]}
     {E_{c}+\tau ln(T_{c}/J_{z})},	
\label{TcHorovitz}
\end{equation}
where $\tau =\phi_{0}^{2}/4\pi e^{2}$ is connected with the BKT
transition temperature as $\tau =8T_{BKT}$.

The comparison of the expressions (\ref{freeenergy}) and (\ref{Omegakin}),
gives the next self-consistent equation 
for the critical temperature $T_{c}$, which follows from (\ref{TcHorovitz}):
\begin{equation}
T_{c}\simeq 4\pi J_{\parallel}
\frac{E_{c}+(\pi J_{\parallel}/2)ln(T_{c}/J_{z})}
     {E_{c}+(4\pi J_{\parallel})ln(T_{c}/J_{z})},	
\label{Tc}
\end{equation}
where the in-plane correlation length
$\xi_{0}$ is absorbed in the parameter $J_{z}$ 
(i.e. $t_{z}(a/\xi_{0})^{2}\rightarrow t_{z}$).
The parameter $E_{c}$ actually should be renormalized by including
the influence of the inter-layer coupling on the vortex system
\cite{Horovitz2}.
It is considered here as a model parameter, which should
be found experimentally, in particular its doping dependence should be taken
into account. For calculation we use the value $E_{c}=0.01W$
(for estimation of $E_{c}$ based on an amplitude dependent
Ginzburg-Landau theory, see for example \cite{Minnhagen}).

It is interesting to note, that in the limit of very small carrier
densities, when $J_{\parallel}\simeq e_{F}\rightarrow 0$,
 the analytical solution for $T_{c}$ can be obtained
$T_{c}\simeq 4\pi J_{\parallel}\simeq e_{F}$.
This is different from the one layer case when
$T_{c}=T_{BKT}\simeq e_{F}/8$, independently on the
pair angular momentum $l$. 
However, the region of extremely low carrier densities is not interesting
from physical point of view.

To find the critical temperature $T_{c}$ one needs to solve
the system of equations 
(\ref{gap}), (\ref{number}) and
(\ref{Tc}) with functions $J_{\parallel}(\mu ,T,\Delta (T))$ and 
$J_{z}(\mu ,T,\Delta (T))$ defined in (\ref{Jpar}) and (\ref{Jperp}).
The numerical solutions
show that $T_{c}<T_{BKT}$ at small carrier densities in the case of 
large values of the inter-layer hopping $t_{z}$ and  not
very strong coupling $G$ (Fig.4).
\begin{figure}[h]
\centering{
\includegraphics[width=8cm,angle=270]{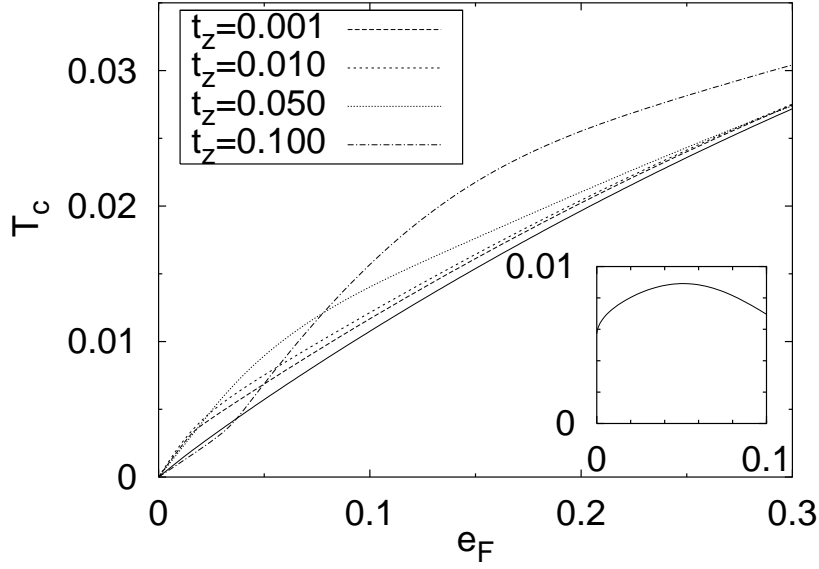}}
\caption{The doping dependence of $T_{c}$
of the layered system in the case $l\not= 0$ at different values 
of the inter-layer hopping
and $G=1.0, E_{c}=0.01$.
The solid line is the corresponding 2D temperature $T_{BKT}$.
The insert is the inter-layer hopping dependence of $T_{c}$
at $G=1$ and $e_{F}=0.05$.}  
\label{fig:4}
\end{figure}
It means that the dependence of the parameters $J_{\parallel}$
and $J_{\perp}$ on coupling, carrier density and temperature
leads to the non-trivial relation between $T_{c}$ and the 2D critical
temperature $T_{BKT}$ at some values of model parameters, different
from $T_{c}>T_{BKT}$, as it was predicted for the case of fixed
$J_{\parallel}$ and $J_{\perp}$ \cite{Hikami,Korshunov,Horovitz1,Horovitz2}.

\begin{figure}[h]
\centering{
\includegraphics[width=8cm,angle=270]{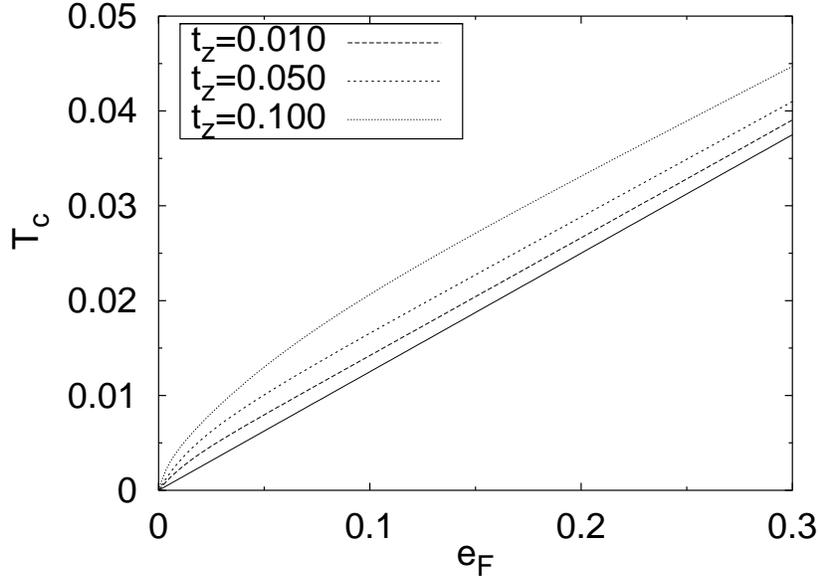}}
\caption{The same as in Fig.4 for the case $l=0$.}  
\label{fig:5}
\end{figure}
In general, $T_{c}$ is growing with  the inter-layer
coupling $t_{z}$ (Figs.4,5).
However, in the case of small carrier
density the critical temperature is decreasing with $t_{z}$ growth
when $l\not= 0$ (Fig.4, in the $l=0$ case this effect takes place
at smaller coupling $G$). 
It can be explained as a consequence of the fact,
that the density of states on the Fermi level $\rho (e_{F})$
at small carrier
densities is decreasing when system tends to become three dimensional
with $t_{z}$ growth ($\rho (e_{F})\simeq \sqrt{e_{F}}$
in the 3D case and $\rho (e_{F})=const$
in the 2D case).
On the other hand, the role of the term $\sim t_{z}^{2}$
must be studied in addition at rather large values of $t_{z}$,
when inter-layer hopping becomes of order of the intra-layer hopping,
i.e. $t_{z}\simeq 0.1W$ (see again \cite{Kim}).

\section{Conclusions}

To summarize, the doping dependence of the superconducting 
critical temperature of layered superconductors 
on the charge carrier density has been studied
in cases of different angular momentum of the pairs $l$, coupling and
inter-layer hopping.
It has been shown that the critical temperature
$T_{c}$ is smaller then the 2D critical temperature $T_{BKT}$
at some values of the model parameters, contrary to the XY-model
with the parameters $J_{\parallel}$ and $J_{\perp}$ which
do not depend on carrier density $n_{f}$, inter-particle coupling $V$
and the temperature of the system $T$.
In particular, 
at small carrier densities $T_c\not= e_{F}/8$,
contrary to the dependence of $T_{BKT}$ in the 2D case.
The critical temperature $T_{c}$ is growing with $t_{z}$, except
the case of  non-zero angular momentum of the pairs
at small carrier densities. 

At the same time some questions are remained unresolved. In particular,
the behavior of the system when the inter-layer coupling $t_{z}$
is not very small has to be studied and  
the doping dependence of the vortex condensation energy
should be taken into account. This problems are planned to be studied 
in the future. 

\section*{Acknowledgments}

V.M.L. acknowledges partial support by SCOPES-project 7UKPJ062150.00/1 
of the Swiss National Science Foundation.

\end{document}